\documentclass{elsart}

\usepackage{epsfig}

\usepackage{amssymb}

\newcommand{\BE}{\begin{equation}}
\newcommand{\EE}{\end{equation}}
\newcommand{\BA}{\begin{eqnarray}}
\newcommand{\EA}{\end{eqnarray}}

\begin{document}

\begin{frontmatter}



\title{Clone size distributions in networks of genetic similarity}

\author[label1]{E. Hern{\'a}ndez-Garc{\'\i}a\corauthref{cor1}},
\ead{emilio@imedea.uib.es}
\author[label1]{A. F. Rozenfeld},
\author[label1]{V. M. Egu{\'\i}luz},
\author[label2]{S. Arnaud-Haond},
\author[label1]{C. M. Duarte}
\corauth[cor1]{Corresponding author. Fax: +34 971 173426}

\address[label1]{Instituto Mediterr\'aneo de Estudios Avanzados
(IMEDEA) \\
CSIC - Universitat de les Illes Balears\\
E-07122 Palma de Mallorca, Spain.}
\address[label2]{CCMAR, CIMAR-Laborat{\'o}rio Associado,
Universidade do Algarve, Gambelas, 8005-139, Faro, Portugal}


\begin{abstract}
We build networks of genetic similarity in which the nodes are
organisms sampled from biological populations. The procedure is
illustrated by constructing networks from genetic data of a marine
clonal plant. An important feature in the networks is the presence
of clone subgraphs, i.e. sets of organisms with identical genotype
forming clones. As a first step to understand the dynamics that
has shaped these networks, we point up a relationship between a
particular degree distribution and the clone size distribution in
the populations. We construct a dynamical model for the population
dynamics, focussing on the dynamics of the clones, and solve it
for the required distributions. Scale free and exponentially
decaying forms are obtained depending on parameter values, the
first type being obtained when clonal growth is the dominant
process. Average distributions are dominated by the power law
behavior presented by the fastest replicating populations.
\end{abstract}

\end{frontmatter}


\section{Introduction}
\label{intro}

Biological systems have always been quoted as archetypes of
complexity. Modern network approaches
\cite{Strogatz2001,Albert02,Dorogovtsev02,Newman2003,Boccaletti2006}
have provided useful insight when applied to many of them, ranging
from protein interaction networks \cite{Jeong2001} to food webs
\cite{Williams2002,Dunne2002}. One of the most fundamental
processes through which organisms interact is that of genetic
interactions, which conforms the basis for evolutionary processes.
The need to represent such processes in network structures more
complex than simple trees is beginning to be appreciated
\cite{Posada2001}, but modern network paradigms are not yet widely
used to understand genetic relationships among individuals,
communities or species.

In this paper we overview the construction of {\sl networks of
genetic similarity}. They are useful tools to represent and
analyze the genetic structure of the different genotypes in a
biological population.
We use the method to organize genetic data from a Mediterranean
marine plant, {\sl Posidonia oceanica}. Then, we move from a
static depiction of these networks to the exploration of the
dynamics that has shaped them. A network characteristic that turns
out to be accessible to mathematical analysis is the degree
distribution of a particular case of similarity network. Such
particular degree distribution is directly related to the clone
size distribution in the population. We construct a dynamical
model for the population dynamics and solve it for the required
distributions. Scale free and exponentially decaying forms are
obtained depending on parameter values. Averages over several
populations, however, are dominated by the power-law behavior
displayed by the fastest clonally growing populations.

\section{Networks of genetic similarity}
\label{sec:networks}

We illustrate our approach with genetic data obtained by
genotyping specimens of {\sl Posidonia oceanica}, a marine plant
living in the coastal waters of the Mediterranean sea
\cite{Larkum2006}. An important characteristic of this organism is
that it is a {\sl clonal plant}, meaning that in addition to the
standard sexual reproduction involving the flowers in different
shoots, the plant also reproduces by growing new shoots or ramets
that are genetically identical to the existing ones. The process
is called {\sl clonal reproduction}, and the set of ramets
generated in this way starting from an initial one form a single
{\sl clone}. Seeds generated by sexual reproduction start new
clones. At short distances, the members of a clone are physically
connected by a common rhizome. Physical connection between distant
ramets can be interrupted by a number of reasons (for example the
death of a part of the intermediate rhizome). From our genetic
data we will still identify them as members of the same clone
because they share the same genotype. Mutations, however, can
occur during the process of clonation. It is customary to consider
these mutants with slightly different genotypes, and the lineages
arising clonally from them, as part of the same original clone. In
this paper, however, we find convenient to take the operational
definition of a clone as the set of ramets with identical
genotype. Within this point of view, mutants are considered to be
founders of new clones.

Our data set comes from about 40 ramets sampled from each of 38
populations across the Mediterranean. For each ramet, the genetic
data consist in the number of microsatellite repetitions at seven
loci. Microsatellites \cite{Goldstein1997} are highly variable
portions of the genome commonly used for intraspecific studies.
They contain a motif consisting of a short sequence of bases which
is repeated for a variable number of times in genetically
different individuals. Additional details on the genetic dataset
can be found in Refs. \cite{Alberto2003,Arnaud2005}. In Ref.
\cite{step1} we introduced a suitable measure of genetic
dissimilarity among the ramets in the data set. This {\sl genetic
distance} between a pair of ramets is defined as the difference
(in absolute value) in the number of motif repetitions present at
a particular genome locus of the two ramets, summed over the seven
analyzed loci. In this way genetically identical individuals
(belonging to the same clone) are at zero distance, and biological
processes can be associated with different distances; for example
there is a characteristic mean genetic distance between parents
and offspring generated by sexual reproduction \cite{step1}.

\begin{center}
\begin{figure}
\epsfig{file=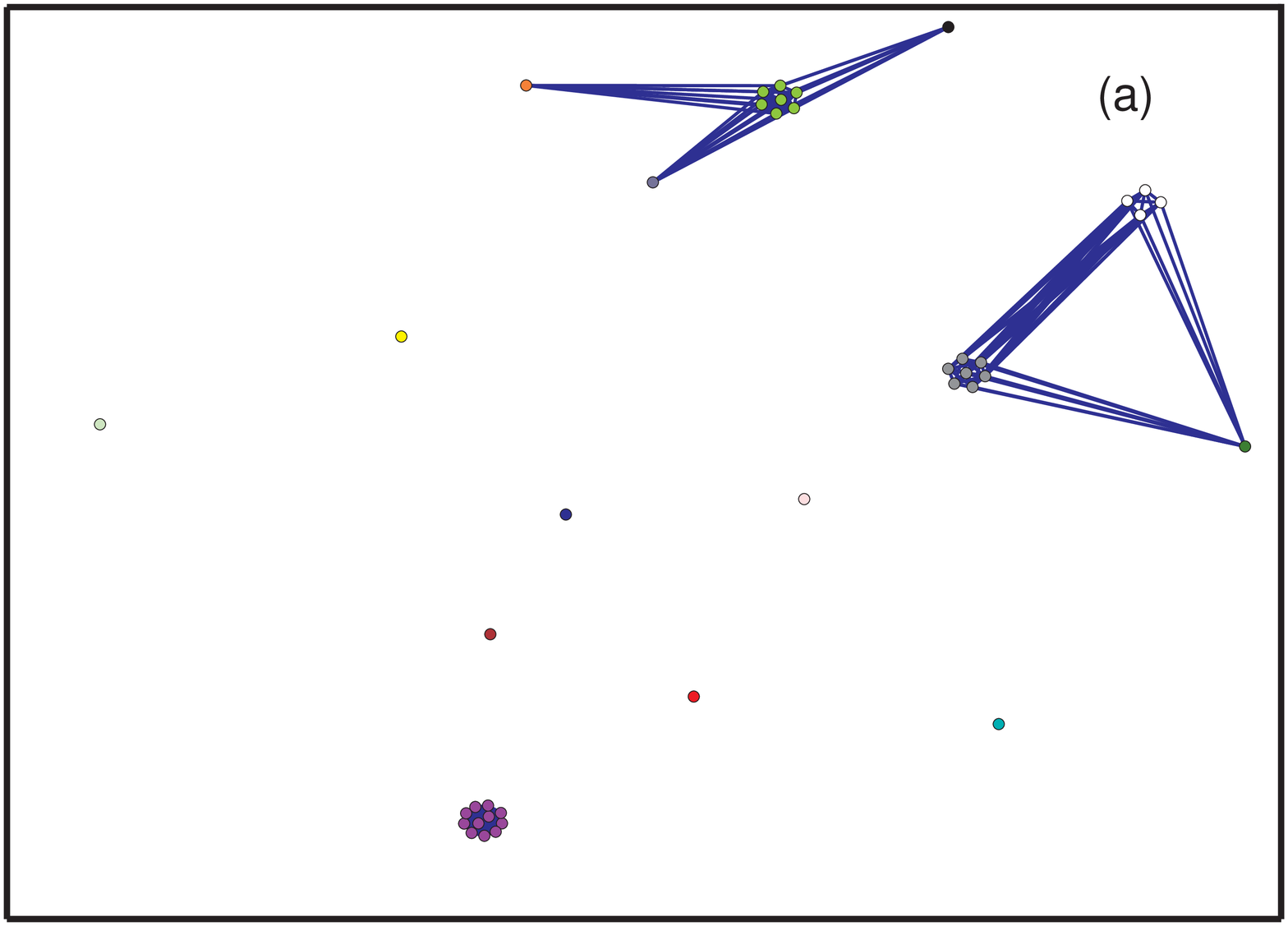,width=0.5\linewidth}
\epsfig{file=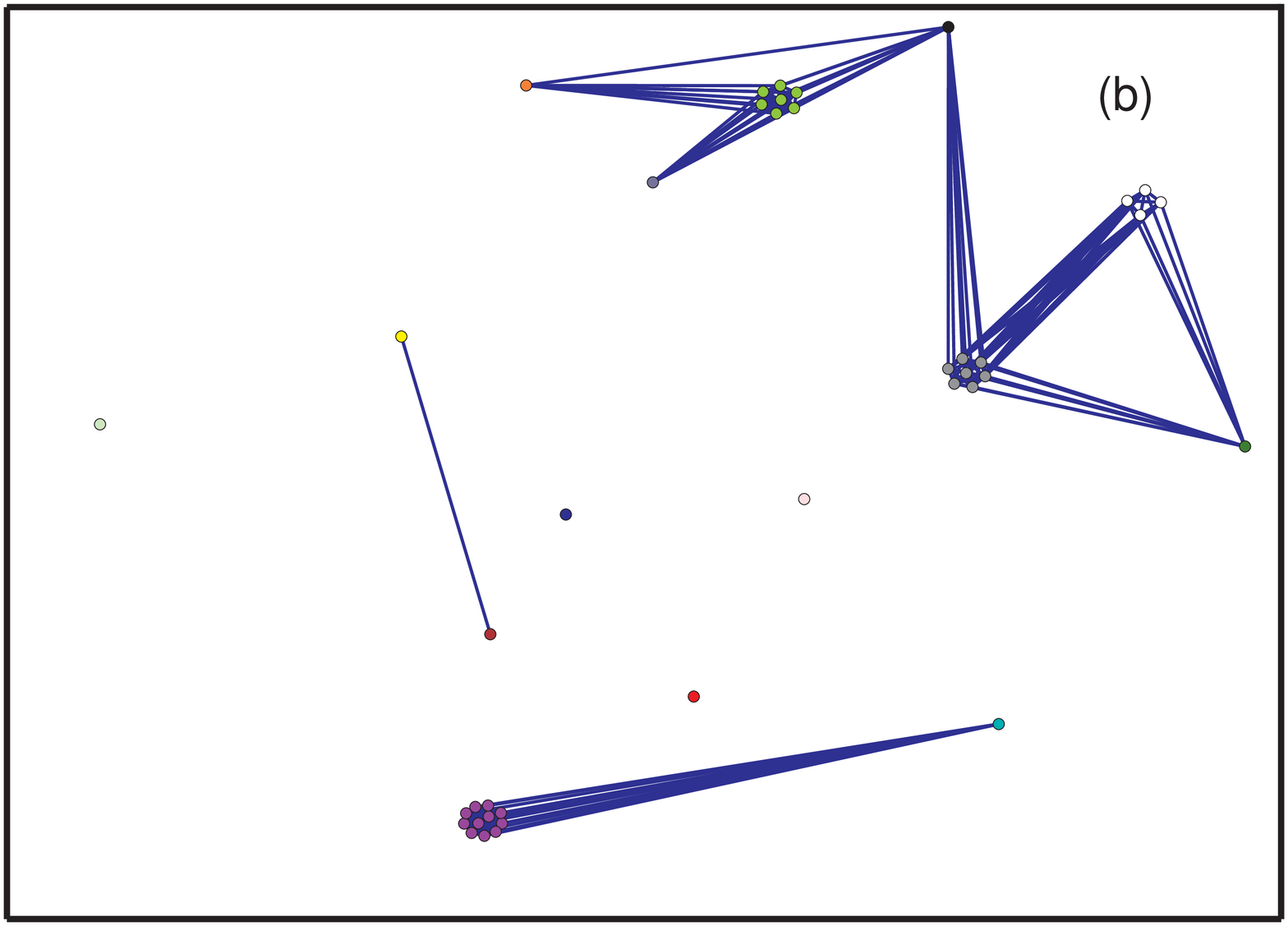,width=0.5\linewidth}
\epsfig{file=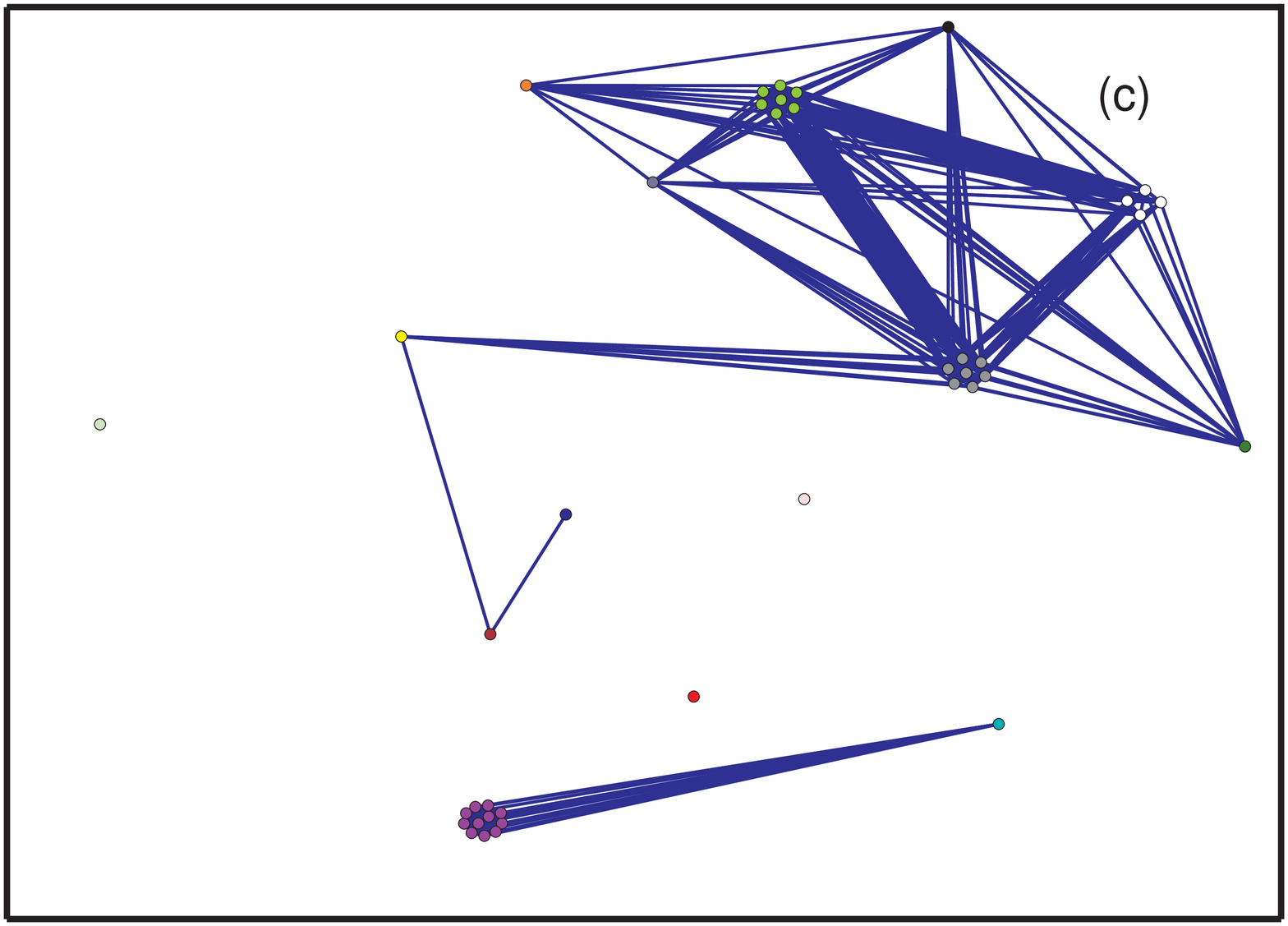,width=0.5\linewidth}
\epsfig{file=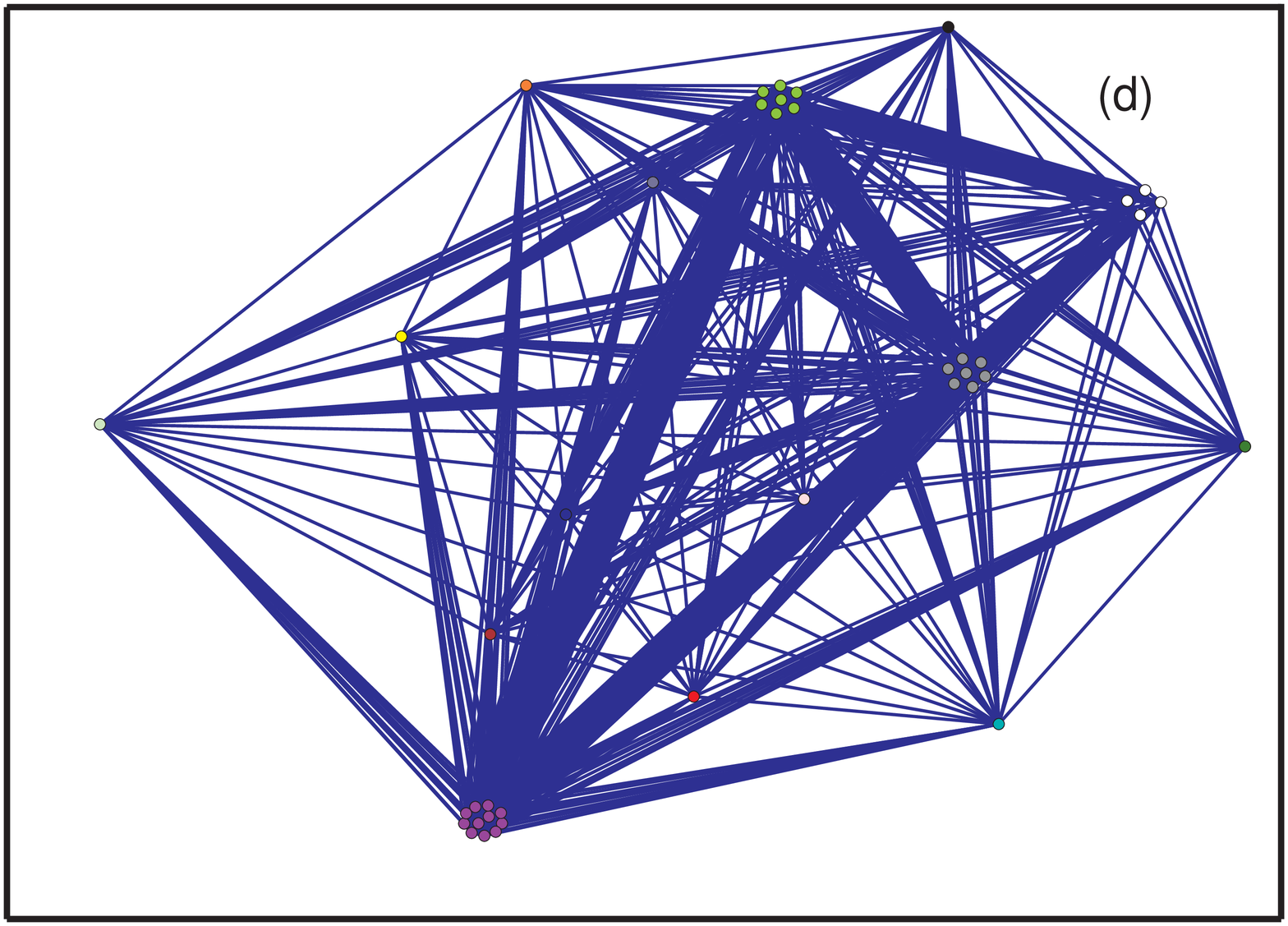,width=0.5\linewidth}
\caption{\scriptsize Similarity networks of ramets from es Cal{\'o} de s'Oli
(island of Formentera, Spain), at different similarity threshold
values: a) 20, b) 27, c) 34, and d) $>$92 (which leads to a fully
connected network). Note the groups of nodes, plotted close
together, which are fully connected subgraphs and remain fully
connected for all values of the threshold. They are the clones.}
\label{fig:nets}
\end{figure}
\end{center}

Once a matrix of genetic distances has been constructed, one can
in principle apply the tools of classical phylogenetics
\cite{Felsenstein2003} to analyze genetic and evolutionary
relationships among the genotypes. Nevertheless, the fact that we
are dealing with intraspecific data and the coexistence of sexual
and clonal modes of reproduction in Posidonia challenges the
applicability of these methods. Aiming at addressing these issues,
networks of genetic similarity have been introduced in
\cite{step1}. In the same spirit as in correlation networks
\cite{Onnela2003,Eguiluz2005,microarrays}, a {\sl similarity
threshold} is introduced in which the individual ramets are
represented by nodes and there is a link between them if their
distance is smaller than the threshold, i.e. if they are more
similar than the threshold value. This gives a sequence of
networks, one for each threshold value. Figure \ref{fig:nets}
displays examples of similarity networks at different threshold
values for the ramets sampled from a population in Es Cal{\'o} de
s'Oli (Formentera island, Spain). Connectivity increases fast
around a threshold value of 30, which is identified in this
particular population with the mean distance arising between
parents and sexual offspring (outcrossing distance).

\begin{center}
\begin{figure}
\epsfig{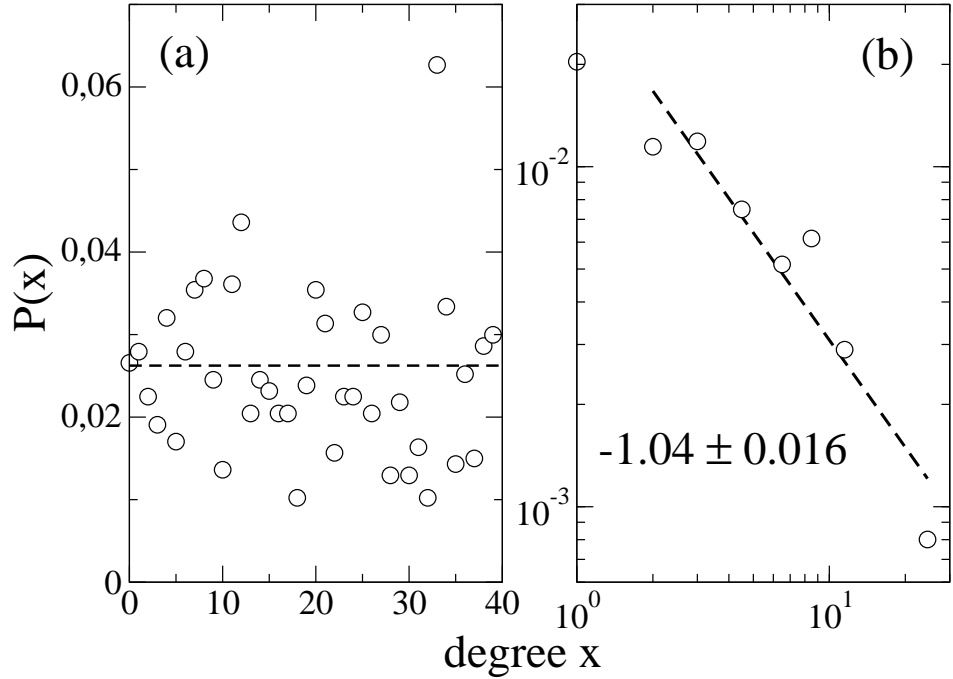}
\caption{\scriptsize Degree distributions, averaged over all the
sampled populations. a) For networks at threshold 30, which is the
most abundant non zero distance in the data set and of the order
of the mean outcrossing distance. Distribution is roughly uniform.
b) Zero threshold, so that only clones remain connected. Data have
been binned by the {\sl data threshold method} as presented in
Ref. \cite{Adami2002}. The distribution is fitted by a power law
with the exponent indicated. }
\label{fig:degree}
\end{figure}
\end{center}

A standard characterization of the topology of a network is the
connectivity degree distribution $P(x)$. It is the proportion of
nodes in a network with a given number $x$ of links. Because of
the limited amount of data, the networks are rather small and the
degree distributions from each of them are noisy and highly
variable. Patterns are clearer when averaging the degree
distributions of the 38 sampled populations. Figure
\ref{fig:degree} shows this averaged distribution at two different
values of the threshold. When the threshold takes a value of the
order of the mean outcrossing distance (Fig. \ref{fig:degree}a),
the degree distribution is essentially flat, although large
statistical fluctuations still remain. To our knowledge, this kind
of flat degree distribution has only been reported previously in
the context of food webs \cite{Dunne2002} (displayed there as
accumulated distributions which vary linearly).

At threshold zero only identical ramets, i.e. pertaining to the
same clone, remain linked. Fig. \ref{fig:degree}b shows the
averaged degree distribution in this case. The amount and range of
the data are small, but the power-law fit (suggested by the
theoretical arguments discussed in the following) indicates that
they are compatible with a degree distribution of the {\sl
scale-free} type. It turns out that, for this zero-threshold case,
the shape of the distribution can be understood in terms of a
model for the population dynamics. The crucial point is to
recognize that there is a relationship between the degree
distribution at zero threshold $P_0(x)$,  and the distribution of
clone sizes, i.e., the number $H_K$ of clones having a number of
ramets $K$. This arises from the fact that clones are fully
connected subgraphs, so that in a clone of $K$ ramets, each ramet
has $K-1$ links. Introducing $M(x) \equiv N P_0(x)$ as the number
of nodes with $x$ links in a network of $N$ ramets, the
relationship is
\BE
H_K= \frac{1}{K}M(K-1)
\label{relation}
\EE
Thus the topological problem of calculating this degree
distribution becomes a question in population dynamics: the
calculation of clone sizes. This last question falls within the
general category of `growth-death-innovation' processes related to
the models of Yule \cite{Yule1925}, Simon \cite{Simon1955} and
others, as summarized in Ref. \cite{Simkin2006}. We devote the
rest of the paper to build and solve a stochastic model of that
type from which the clone size distribution and other useful
quantities can be analytically extracted. We stress that the
topological structure of the network at similarity threshold zero
is very simple: it consists of disjoints sets (the clones) of
nodes which remain internally fully connected. Thus, standard
network descriptors have rather trivial values in this case (for
example the clustering coefficient is always one, independently of
degree). This is why we focus on the degree distribution, and in
the number of connected components (given by the richness defined
below), as the only nontrivial quantities characterizing the
network at vanishing threshold.

\section{Modelling population and clone dynamics}
\label{sec:model}

A plant population (a meadow) consists of a variable number
$N=N(t)$ of ramets. We label the different genotypes in the
population --the different clones-- with numbers $j$,
$j=1,2,...,G$, so that $G$ is the richness of the population, i.e.
the number of different clones. $K_j$ is the number of ramets
forming clone $j$. We have $N =\sum_{j=1}^{G} K_j$. The histogram
$H_K=\sum_{j=1}^{G} \delta_{K,K_j}$ counts how many clones of size
$K$ are in the population. We have
\BE
G=\sum_{K=1}^{\infty} H_K \ ,\
\label{GofH}
\EE
and
\BE
N=\sum_{K=1}^{\infty} K H_K
\label{NofH}
\EE

As a major simplification in our approach, we neglect any
difference among the ramets arising from their age, genetic
content, or integration into big or small clones. Thus all ramets
are described by the same parameters of mortality, reproduction,
etc. Additionally we consider a well mixed population so that
spatial effects are not taken into account. The local density is
the same everywhere and simply proportional to the population size
$N$. Finally, we do not take into account any possible seasonal
modulation of the different rates. The limitations introduced in
our modelling approach by these simplifications will be briefly
discussed in Section \ref{sec:discus}.

We now introduce the elementary processes occurring in our
stochastic model (mortality, clonal and sexual reproduction,
mutation, and migrations). They are considered to be independent
events occurring at Poisson times:

\begin{itemize}
\item[1.-] {\sl Death at rate $d$}. Each ramet has a probability $d$ per
unit of time of dying. In {\sl Posidonia oceanica} mortality is
not strongly dependent on population density, so that we could
take its value to be a constant. But in order to keep the model
general enough we include some competition for resources in terms
of a mortality that increases with population size:
$d=d(N)=d_0+\nu N$.

\item[2.-] {\sl Clonal reproduction at rate $c$}. The quantity $c$ is
the probability of ramet clonal reproduction per unit of time,
which may be density dependent: $c=c(N)=\max\left[0, c_0\left(
1-N/N_s \right)\right]$. $N_s$ is the number of ramets in the
population beyond which clonal reproduction becomes impossible.
This saturation effect is relevant in {\sl Posidonia} populations,
and fixes a maximum density for the meadow. The max function is
needed to avoid negative probabilities. If the dynamics keeps $N$
sufficiently below $N_s$ we can approximate $c \approx c_0\left(
1-N/N_s\right)$.

\item[3.-] {\sl Mutations with probability $p_m$}. In the process
of clonal reproduction a mutation can occur with probability
$p_m$. We assume that each mutant genotype is different from the
ones already present in the population, and thus founds a new
clone. We neglect the possibility of mutation backwards to the
original genotype.


\item[4.-] {\sl Sexual reproduction at rate $s$}. At first
sight, the number of births arising from sexual reproduction
should be proportional to the product of the number of male and
female flowers in the population, and thus roughly to $N^2$. But
in {\sl Posidonia} populations, as well as in other plant species,
pollen is produced in great abundance so that reproduction is
rather proportional to the number of females. Since Posidonia
flowers are hermaphrodite, this equals the total number of
flowers, that we assume to be proportional to the number of ramets
$N$. Nevertheless, we keep the modelling at a general level and
include both a linear and a quadratic dependence on $N$ for the
number of sexual births, which implies a sexual reproduction rate
per ramet of $s=s(N)=s_0+\epsilon N$. Sexual reproduction produces
unique individuals as a consequence of the random recombination of
the progenitor's genomes. Thus, each newborn founds a new clone.
Sexual reproduction in plants is strongly seasonal, but we
consider here the process as occurring continuously in time.

\item[5.-] {\sl Migration, $I$}. Migration process (transport of
seeds or broken ramets by
marine currents, animals or ships) will only renormalize the death
parameters $d_0$ and $\nu$ when they imply a loss from the
population. These changes will not be explicitly written down.
When there is input of ramets from outside the population we
assume that immigrants are genetically distinct from any of the
local clones, so that they also found new clones on arrival. We
will call $I$ the number of immigrants entering the population per
unit of time.

\end{itemize}

Summarizing the modelling parameters, we have the basic death,
clonation, and sexual reproduction rates of ramets at low
densities ($d_0$, $c_0$, and $s_0$, respectively), the
modifications to these rates from interactions ($\nu$, $N_s$, and
$\epsilon$, respectively), the probability of mutation while
cloning ($p_m$), and the immigration rate ($I$). There are strong
differences among different populations, leading to rather
different network structures \cite{step1}. This reflects the fact
that they are in very different states, some of them healthy, many
of them receding, and are subjected to a large variety of
environmental parameters and pressures. Accordingly, we will
postpone the discussion on the election of numerical values for
the different parameters until Section \ref{sec:discus}, in which
we will argue that {\sl average} quantities such as the
distributions plotted in Fig. \ref{fig:degree} are dominated by
populations having particular parameter values. Until then we keep
the discussion general, using arbitrary parameter values, which is
also convenient for developing a formalism that can be later
applied to species different from the one considered here.

\subsection{Population size rate equation}
\label{subsec:poprate}

It is simple to write down the rate equation for the time
evolution of the expected population size $N$. Neglecting all
fluctuations and correlations, and taking into account that
mutations do not alter the number of individuals, the net growth
rate per ramet is $\gamma_T=c(N) + s(N) -d(N)$, so that:
\BE
\dot N = \left( c(N) + s(N) -d(N) \right) N + I = \beta N \left(
1-\frac{N}{N_c}\right)  + I \ ,
\label{popequation}
\EE
where $\beta \equiv \left( c_0 + s_0 - d_0 \right)$ is the maximum
population growth rate, and we have introduced the carrying
capacity $N_c$ from the relation
\BE
\frac{\beta}{N_c} \equiv \frac{c_0}{N_s}+\nu-\epsilon
\EE
or
\BE
N_c = \frac{\beta N_s}{c_0+N_s(\nu-\epsilon)}  \ .
\EE

In the absence of migration, $I=0$, Eq. (\ref{popequation}) is
just the logistic equation. If $\beta>0$ any initial condition
approaches the saturation value $N_c$. Solutions from small
initial values of $N$ have an initial exponentially growing phase
($N\approx N_0 e^{\beta t}$). If $\beta<0$ the solution decays
towards zero (extinction), exponentially if $N$ is sufficiently
smaller than $N_c$. We notice however that stochastic
fluctuations, neglected in Eq. (\ref{popequation}), will be
important when the population is small.

\subsection{Expected clone size}
\label{subsec:clonesize}

Let us focus on clone $j$ of size $K_j = K_j(t)$ (again we will
neglect its statistical fluctuations). Sexual reproduction,
mutations and immigrants do not change it. Thus
\BE
\dot K_j(t) = \gamma_c K_j(t) ,
\label{clonsize}
\EE
with $\gamma_c(N) \equiv  g_c(N)-d(N)$. We call $g_c(N) \equiv
(1-p_m)c(N)$ the clonal growth rate, and $\gamma_c$ is a net
clonal growth rate.

On average, sexual reproduction, mutations and immigration
increase the richness $G$ by starting an amount $\left( p_m c(N) +
s(N) \right) N+I$ of new clones of size 1 per unit of time. As a
consistency check, one can see that the sum of Eq.
(\ref{clonsize}) over all clones ($\sum_{j=1}^G K_j(t)=N(t)$),
with the addition of the new ones, leads to the population
equation (\ref{popequation}).

Once $N(t)$ is obtained from (\ref{popequation}), Eq.
(\ref{clonsize}) can be integrated exactly. The simplest situation
corresponds to $\nu=\epsilon=c_0/N_s=0$, so that ramets do not
interact (in this case Eq. (\ref{clonsize}) is exact for the
average clone size even in the presence of fluctuations). Then,
$\gamma_c$ is a constant in time ($\gamma_c=(1-p_m)c_0-d_0$), and
$K_j(t)=K_j(0)e^{\gamma_c t}$. We note that $\gamma_c<\gamma_T$,
so that clones will only grow on average when the total population
is growing fast enough. It is interesting to note that the total
population can be growing while on average each existing clone is
decreasing in size. A similar situation occurs for interacting
ramets when $N(t)$ is constant in time (this happens when $N$ has
attained its saturation value, which in the absence of immigration
is $N=N_c$). In this case $K_j(t)=K_j(0)e^{\gamma_c t}$, with
$\gamma_c=\gamma_c(N_c)<0$. On average, existing clones will
decrease in size until dying. In the mean time, mutations, sex and
immigration will introduce clones of size one (single ramets), and
this together with stochastic fluctuations will maintain a steady
state distribution of clone sizes peaked at small values.

\subsection{Clone size distribution}
\label{subsec:clonedist}

The next step is to estimate the whole clone size distribution
$H_K$ in the population. By balancing the different rates, we find
that the expected value of that distribution is ruled by:
\BA
\frac{d}{dt} H_K(t) &=& g_c \left( (K-1)H_{K-1}(t)-K H_K(t)
\right) \nonumber \\
 &+& d \left( (K+1)H_{K+1}(t)-K H_K(t) \right) \ ,\ \ \ K>1 \nonumber \\
\label{Hk}  \\
 \frac{d}{dt} H_{1}(t) &=& I + h N(t) - g_c H_{1}(t) + d \left(
2H_{2}(t)- H_{1}(t) \right) \ .
 \label{H1}
\EA
This mean field description becomes exact when ramets do not
interact. The increase in $H_{1}(t)$, the number of clones
consisting of a single ramet, arises from immigration $I$ and from
mutants and sexual offspring: $h=h(N) \equiv p_m c(N) +
s(N)=\gamma_T-\gamma_c$. This quantity $h$ can be thought as the
amount of {\sl innovation}, since it gives the rate at which new
genotypes are produced by the system. Note that $N(t)$ in Eq.
(\ref{H1}) is not an independent variable, but it is related to
all the $H$'s by Eq. (\ref{NofH}). Thus the single-ramet
population $H_1$ plays a special role since it interacts with all
the clone-size groups. Equation (\ref{popequation}), which is
closed for $N(t)$, is recovered from Eqs. (\ref{NofH}) and
(\ref{Hk})-(\ref{H1}). If one uses first this equation to find
$N(t)$, one of the equations in (\ref{Hk})-(\ref{H1}) becomes
redundant.

From Eqs. (\ref{Hk})-(\ref{H1}) one can obtain an equation for the
richness (\ref{GofH}):
\BE
\frac{d}{dt} G(t) = I+h N(t)-d H_1(t) \ .
\label{richness}
\EE
%

We analyze now Eqs. (\ref{Hk})-(\ref{H1}) in some particular
cases. For simplicity we neglect immigration: $I=0$, and we only
address the situations of noninteracting ramets
($\nu=c_0/N_s=\epsilon=0$), and of interacting ramets in a
constant population $N(t)=N_c$. In these cases, (\ref{Hk}) are
effectively linear equations with coefficients constant in time,
and the explicit time dependence of $N(t)$ in (\ref{H1}) can be
written as $N(t)=N_0 e^{\gamma_T t}$. In the noninteracting case
$\gamma_T=\beta$, $h=p_m c_0+s_0$, and we can have either a
growing population if $\beta=c_0+s_0-d_0>0$ or a decaying one if
$\beta<0$. Note that this description also applies to the
interacting ramet case when the population is small enough,
because the interaction terms become negligible when $N\approx0$.
In the interacting constant population case $\gamma_T=0=h+g_c-d$
and $N_0=N_c$

In all cases we search for solutions of the type
\BE
H_K(t)=F_K e^{rt} \ .
\label{HofF}
\EE
We see from substitution in Eq. (\ref{NofH}) that
\BE
N(t)=e^{rt}\sum_{K=1}^\infty K F_K \equiv e^{rt} R
\label{NofF}
\EE
so that necessarily $r=\gamma_T$. The sum in Eq. (\ref{NofF}),
which we call $R$, should converge to $N_0$ or $N_c$.

Substituting Eq. (\ref{HofF}) into (\ref{Hk})-(\ref{H1}) we get
\BA
 \gamma_T F_K &=& g_c (K-1)F_{K-1}+ d (K+1)F_{K+1}  - (g_c+d) K F_K \ \ , \ \ \ \ K>1
\label{Fk}  \\
 \gamma_T F_{1} &=&  h R - (g_c+d) F_{1} + 2 d F_{2}
 \label{F1}
\EA
Equation (\ref{Fk}) is a second order linear recurrence, having in
principle two independent solutions. The point is that only one of
them will satisfy the constraint (\ref{NofF}) with a convergent
sum $R$. Fixing $N_0$ or $N_c$ ($=R$) then normalizes and
completely determines the distribution.

We find the asymptotic behavior for large $K$ of the two possible
solutions of Eq. (\ref{Fk}) by substituting the ansatz
\BE
F_K \sim \frac{\theta^K}{K^z} \ .
\label{FofK}
\EE
The first two terms in an expansion in powers of $K^{-1}$
determine two independent solutions:
\BA
A: \hskip 1cm &\theta=1 \ ,\ \    &z=1+\frac{\gamma_T}{\gamma_c}=2+\frac{h}{\gamma_c}  \label{solA}  \\
B: \hskip 1cm &\theta=\frac{g_c}{d} \ ,\ \
&z=1-\frac{\gamma_T}{\gamma_c}=-\frac{h}{\gamma_c}
\label{solB}
\EA

\begin{center}
\begin{figure}
\epsfig{file=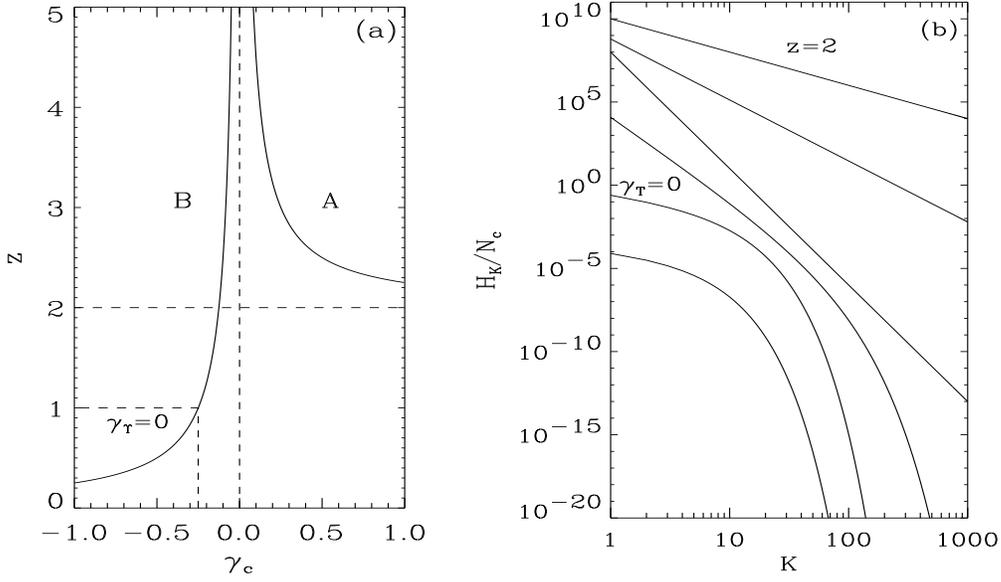,width=0.6\linewidth,height=1.\linewidth,angle=270}
\caption{\scriptsize a) Exponent z as a function of net clonal growth $\gamma_c$.
$d=1$ and $h=0.25$.  The place where $\gamma_T=0$, giving $z=1$,
is indicated. b) Clone distributions. $d=1$, $h=0.25$, and
$\gamma_c$ increasing from top to bottom: $g_c=0.6, 0.75, 0.95,
1.05, 1.15$. The upper curve is the limit case $g_c >> h$, leading
to $z=2$. The line labelled $\gamma_T=0$ ($g_c=0.75$) is the exact
solution given by Eq. (\ref{logarithmic}), and the other curves
have the proper asymptotic behavior but have been arbitrarily
shifted vertically for clarity. For $\gamma_c>0$ (three upper
lines) solution A (Eq. (\ref{solA})) gives a power law behavior
with exponent $z$. For $\gamma_c<0$ (three lower curves) the
exponent in solution B (Eq. \ref{solB})) gives only a
pre-asymptotic power law behavior, being the final decay of
exponential type.}
\label{fig:distributions}
\end{figure}
\end{center}

The values of the exponent $z$ depend on the ratio between the net
growth $\gamma_T$ (or of the innovation $h$) and the clonal net
growth $\gamma_c=g_c-d$. We analyze separately the following three
cases: $\gamma_T=0$, $\gamma_c<0$ (which includes cases with
$\gamma_T>0$ and $\gamma_T<0$), and $\gamma_c>0$ (which only
happens when $\gamma_T>0$). When $\gamma_c=0$ the ansatz
(\ref{FofK}) is not appropriate.

\begin{itemize}
\item[1.-] {\sl Constant population size ($\gamma_T=0$)}. Since $h+g_c-d=0$
we have $g_c<d$, and the size of a given clone is decreasing on
average. Solution A (Eq. (\ref{solA})) makes the sum in
(\ref{NofF}) to diverge and B is the correct solution. It turns
out that it gives not only the asymptotic behavior, but an exact
solution for all $K$, the so-called {\sl logarithmic
distribution}:
\BE
H_K(t) = \frac{hN_c}{g_c} \left( \frac{g_c}{d}\right)^K
\frac{1}{K} = N_c \left(\frac{d}{g_c}-1 \right)\left(
\frac{d}{g_c}\right)^{-K} \frac{1}{K} \ .
\label{logarithmic}
\EE
As expected, this is concentrated at small clone sizes, with a
quasi-exponential decay at large $K$. When the mutation and sex
rates are small (so that $g_c$ approaches $d$) the range of
existing clone sizes becomes larger. In \cite{Karev2002} this
exact solution is also found for a related set of equations in the
context of the `simple BDIM model' of protein evolution, and its
stability is proven. Thus, arbitrary initial distribution of
clones will converge at long times towards Eq. (\ref{logarithmic})
when the population has reached a constant value.

The richness can be calculated from (\ref{richness}). The relative
richness $v$ becomes:
\BE
v \equiv \frac{G}{N_c}=\left(1-\frac{d}{g_c} \right)
\log\left(1-\frac{g_c}{d}\right) \ .
\label{richnessNconstant}
\EE
This relative richness is a decreasing function of $g_c/d$, from a
maximum value in the case of maximum mutations and sex
($g_c/d\approx 0$, $v\approx 1$) to a vanishing value in the case
of minimum mutations and sex ($g_c/d\approx 1$, $v\approx 0$).

\item[2.-] {\sl Decreasing average clone size ($\gamma_c=g_c-d<0$)}.
This case includes situations in which population is decreasing
($\gamma_T=\beta<0$), but also cases in which there is net growth
($\gamma_T=\beta>0$) but not enough of the purely clonal type. The
sum in (\ref{NofF}) diverges for solution A (Eq. (\ref{solA})) and
again B (Eq. (\ref{solB})) gives the correct asymptotic behavior.
Thus there are solutions to (\ref{Hk})-(\ref{H1}) that behave at
large $K$ as
\BE
H_K(t) \sim \frac{ e^{\beta t} }{K^z}\left( \frac{g_c}{d}
\right)^K
\label{solg<m}
\EE
with
\BE
z=1-\frac{\beta}{\gamma_c} =1-\frac{c_0-d_0+ s_0}{c_0 - d_0 -p_m
c_0} \ .
\label{z2}
\EE
The asymptotic behavior is again quasi-exponential, with faster
decay for smaller $g_c/d$. As if trying to compensate for this
increased steepness, the exponent of $K$, $-z$, becomes less
negative when decreasing $g_c/d$ (see Fig.
\ref{fig:distributions}a), starting from a divergent value when
$g_c \approx d$, crossing to $-z=-1$ when $\beta=0$ (formally
coinciding with the solution for constant population), and
approaching $z=0$ when $g_c/d \rightarrow 0$. Note that, despite
this asymptotic behavior, there is a pre-asymptotic power law
behavior with exponent $z$, which can be noticeable (see Fig.
\ref{fig:distributions}b, third curve from bottom) when $\gamma_c$
is close to zero.

\item [3.-] {\sl Growing average clone size ($\gamma_c=g_c-d>0$)}. This situation
requires a net population growth ($\gamma_T=\beta>\gamma_c>0$). In
this case solution B (Eq. (\ref{solB})) is not normalizable and
the power law solution A (Eq. (\ref{solA})) is the only acceptable
asymptotic behavior at large $K$:
\BE
H_K(t) \sim \frac{ e^{\beta t} }{K^z}
\label{solg>m}
\EE
with
\BE
z=1+\frac{\beta}{\gamma_c}=1+\frac{c_0-d_0+ s_0}{c_0 - d_0 -p_m
c_0} \ .
\label{z1}
\EE
This result coincides with the one in Refs.
\cite{Zanette2001,Manrubia2002} for family name distributions
obtained from a discrete-time model of growing families. It
reduces to the classical result of Simon \cite{Simon1955},
explaining the Zipf's observations on the growth of cities, if
$d=0$ so that $z=2$. The clone distribution of a growing
population reaches a power law with the exponent related to the
quotient between the net growth and the net clonal growth.
Starting from the less negative exponent, the one with $z=2$, the
power law has a more and more negative exponent when $g_c$
approaches $d$ from above, so that the clone distribution becomes
more concentrated, until reaching $g_c=d$. Numerical simulations
in Refs. \cite{Zanette2001,Manrubia2002} show (within their
discrete model) that the state given by Eq. (\ref{solg>m}) is
actually approached at long times, although long transients can
occur.
\end{itemize}


\section{Discussion and perspectives}
\label{sec:discus}

We have found that, within the hypothesis of our model, clone
distributions decay at large $K$ either as a power law or
exponentially (with power law corrections). The first situation
occurs when clones are growing on average, which requires a global
population growth, and the second when clones shrink on average.

\begin{center}
\begin{figure}
\epsfig{file=CantClonesVSsizeWithInset.eps,width=0.5\linewidth}
\epsfig{file=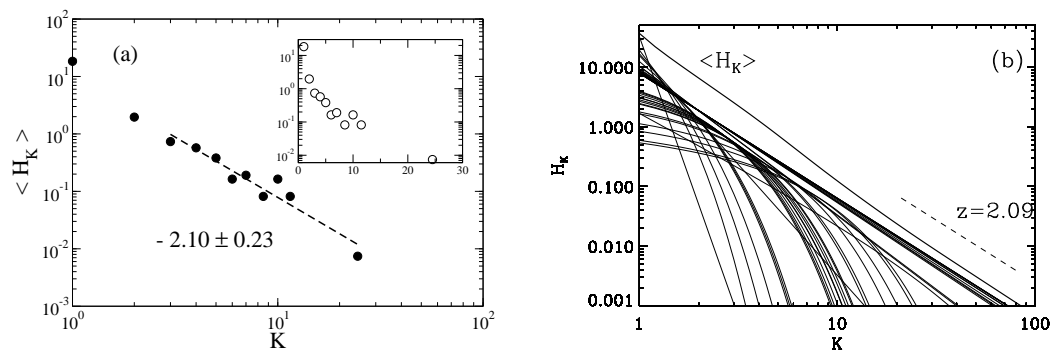,width=0.5\linewidth}
\caption{\scriptsize a) Clone distribution
averaged over the 38 sampled populations. Data are binned as in
Fig. \ref{fig:degree}b, and presented in doubly logarithmic scale
with a power-law fit leading to $z\approx 2.1$. The inset shows
the same data in semilogarithmic scale. b): Thin continuous lines
are 38 distributions $H_K$ of the type A or B, with $d=1$,
$h=0.05$, and random values of $g_c$ as explained in the text.
Thick continuous line is their average $<H_K>$, arbitrarily
displaced vertically for clarity. Its asymptotic behavior at large
$K$, $K^z$, is indicated by a dashed line and is the power law
corresponding to the largest value of $g_c$ in the averaged
sample. In the particular set averaged here, it is $g_c\approx
1.54$, leading to $z \approx 2.09$. }
\label{fig:clonesizes}
\end{figure}
\end{center}

Figure \ref{fig:clonesizes}a displays the observed clone
distribution averaged over all sampled plant populations, obtained
from the degree distribution in Fig. \ref{fig:degree} by using Eq.
(\ref{relation}) (or equivalently by directly counting the clones
in the populations and averaging the resulting histograms). The
asymptotic behavior is consistent with a power law of exponent
$z\approx 2.1$ (for completeness, however, and given the small
range of the data, we show also in the inset a semilogarithmic
plot of the same points which gives only a slightly worse
representation). The appropriateness of the power law fitting
seems to imply that data are better explained by a model in which
clones are growing, which needs population grow, and that there is
a small proportion of sexual reproduction and mutations with
respect to clonal growth.

Although certainly it is realistic to have a rate of accurate
clonal reproduction much larger than sexual or mutational rates,
most of the studied populations were in recession rather than
growing, and some of them in serious danger of disappearance.
Given the limitation in number and range of the data presented in
Fig. \ref{fig:clonesizes}a, we can not exclude that the sampled
clone sizes are not in the asymptotic range addressed by the
theory, or the possibility of functional forms other than power
laws, but the proximity of the fitted slope to the particular
value $z=2$ suggest us the following argument: We believe that
data in Fig. \ref{fig:clonesizes} do not reflect a generally
growing population state, but rather a mixed state in which some
populations are stable or receding, but with the average clone
distribution dominated by a few populations in a clonally growing
state. To check this we plot in Fig. \ref{fig:clonesizes}b a set
of distributions of types (\ref{solg<m}) and (\ref{solg>m}). They
have been generated by choosing units of time such that $d=1$,
fixing $h=0.05$, and randomly generating values of $g_c$ from a
uniform distribution in $[0.0,1.6]$. All the distributions are
normalized to the same value $\sum_1^\infty K H_K=N=40$ as
reflecting that the same number of ramets has been sampled from
each population independently of its true total size. The mean
value of $g_c$ is 0.8, so that the mean values of both the net
growth and the clonal net growth are negative. Nevertheless, the
asymptotic behavior of the average distribution is a power law
dominated by the particular population which turns out to be
growing and having the value of $z$ closest to $z=2$.

The dominance of the populations with the highest clonal net
growth (the {\sl fastest replicators}) would not be restricted to
interpopulation averages: As soon as there is some diversity
inside a population (different clones differ in size, genotype,
location, age, ...) the same arguments imply that the
subpopulation with the highest clonal net growth would dominate
the large size part of the clone distribution in a meadow. We
expect thus a bias in the observed clone distributions towards
power laws characterized by exponents $z$ larger than but close to
$z=2$ (and associated degree distributions $P_0(x)$ behaving
asymptotically as $x^{-q}$, with $q$ larger than but close to
$q=1$). Testing these expectations for Posidonia meadows should
wait until the availability of more abundant sets of
intrapopulation genetic data, to achieve the necessary statistical
power.

Along this paper we have defined clones as sets of genetically
identical ramets. Our results can be directly applied to clones
defined as sets of ramets arising from clonal reproduction, i.e.
considering that both identical ramets and their mutants pertain
to the same clone. Size distributions and the rest of properties
in Eqs. (\ref{logarithmic})-(\ref{z1}) are obtained for these
clones containing several genotypes simply by putting $p_m=0$ in
the corresponding formulae, since in this way mutants are counted
together with the perfect clones instead of with the sexual
newborns. The {\sl innovation parameter} $h$ is reduced and then
the new size distributions are slightly more {\sl heavy tailed}.

We note that, at the modelling level, sexual reproduction enters
just as one more contribution to the innovation parameter $h$. The
same role is played by mutations or any other process generating
new genotypes from old ones. Thus we expect our approach and
results to be applicable to other types of organisms with or
without sexual reproduction capabilities, but having an
appropriate source of innovation such as mutation.

Several simplifying assumptions limit the generality of our model.
Consequences of relaxing the assumption of complete equivalence of
the ramets inside the same population have been mentioned before.
Spatial effects have also been neglected. Although the assumption
of perfect mixing may be justified for the sexual mode of
reproduction in populations of sufficiently small extent, clonal
growth is an inherently local process. It may happen that,
although the mean field density could be small and then not
limiting the growth, ramets could be spatially concentrated and
competition will make growth smaller than expected. In addition,
growth occurs mainly on the {\sl periphery} of the clones
\cite{Sintes2005}. Associated with these spatial effects are the
consequences of statistical fluctuations, enhanced by the discrete
nature of individual ramets and mostly neglected in the present
work. Fluctuations should be important at least close to the
boundary to extinction regimes \cite{Oborny2005}. We believe that
a first consequence of all these effects would be a shift of
effective growth rates towards lower values, but taking them
properly into account would require extensive computer simulation.
An additional complication is that, very likely, any long lived
meadow has suffered periods of growth, periods of enhanced
mortality, etc. The distributions observed presently in the
different populations are probably complicated transient states
resulting from all the past history. The impact of time-dependent
parameters should be addressed to have a complete description of
the processes shaping clone distributions and in general the
genetic structure of the populations. This temporal variability,
together with the consideration of explicit space dependence will
be the subject of future work.

Finally, the focus of this paper has been on the clone subgraphs,
structures with a rather peculiar structure. Dynamical modelling
of the ecological processes shaping the whole network of genetic
similarity of a population remains an open challenge.

\section*{Acknowledgments}

This research was funded by a project of the BBVA Foundation
(Spain), by project NETWORK (POCI/MAR/57342/2004) of the
Portuguese Science Foundation (FCT) and by the project CONOCE2
(FIS2004-00953) of the Spanish MEC. S.A.H. was supported by a
postdoctoral fellowship from FCT and the European Social Fund and
A.F.R. by a post-doctoral fellowship from the Spanish Ministry of
Education and Science.

\end{document}